\documentclass[traditabstract]{aa} 

\usepackage{graphicx}
\usepackage{txfonts}

\newcommand{\myrule}{\rule[-0.1cm]{0.cm}{0.7cm}} 
\newcommand\lsun{L_{\odot}}
\newcommand\msun{M_{\odot}}

\newcommand\mjup{M_\mathrm{Jup}}

\newcommand\chaha{Cha\,H$\alpha$\,}

\begin{document}
	
\title{Improved radial velocity orbit \\ of the young binary brown dwarf candidate \chaha8\thanks{Based 
     on observations obtained at the Very Large Telescope of the 
	    European Southern Observatory at Paranal, Chile 
	    in program 
	    279.C-5018(A),   
	    380.C-0619(A),   
	    082.C-0023(A+B), 
	    385.C-0510(A).   
	  }}

 \titlerunning{Improved radial velocity orbit of the BD/VLM binary \chaha8}

   \author{V. Joergens
          \inst{1,2}
          \and
          A. M\"uller\inst{1,3}
	  \and
	  S. Reffert\inst{4}
          }

   \institute{Max-Planck Institut f\"ur Astronomie, 
     K\"onigstuhl~17, D-69117 Heidelberg, Germany,
             \email{viki@mpia.de}
	 \and
	     Zentrum f\"ur Astronomie Heidelberg, 
	     Institut f\"ur Theoretische Astrophysik,
	     Albert-Ueberle-Str. 2, 69120 Heidelberg, Germany
	 \and
	     European Southern Observatory,
             Karl-Schwarzschild-Str. 2, 85748 Garching, Germany
         \and
             Zentrum f\"ur Astronomie Heidelberg, Landessternwarte, 
             K\"onigstuhl 12, 69117 Heidelberg, Germany
             }

   \date{Received ; Accepted 10 June 2010}

  \abstract
   {The very young brown dwarf candidate \chaha8 was recently discovered
     to have a close ($\sim$1\,AU) companion by means of radial velocity monitoring. 
     We present additional radial velocity data obtained with UVES/VLT between 2007 and 2010,
     which significantly improve the orbit determination of the system.
     The combined data set spans ten years of radial velocity monitoring for \chaha8.
     A Kepler fit to the data yields an orbital period of 
     5.2\,years, an eccentricity of $e$=0.59, and a radial velocity semi-amplitude of 2.4\,km\,s$^{-1}$. 
     A companion mass $M_2\sin i$ (which is a lower limit due to the unknown orbital inclination $i$)
     of 25$\pm$7\,$\mjup$ and of 31$\pm$8\,$\mjup$ is derived when
     using model-dependent mass estimates for the primary of 0.07\,$\msun$ and 0.10\,$\msun$, respectively.
Assuming random orientation of orbits in space, we find a very high probability 
that the companion of \chaha8 is of substellar nature:
With a greater than 50\% probability ($i \geq$ 60$^{\circ}$), the companion mass is between 30 and 35\,$\mjup$
and the mass ratio $M_2$/$M_1$ smaller than 0.4;
with a greater than 87\% probability ($i \geq$ 30$^{\circ}$) the companion mass is 
between 30 and 69\,$\mjup$ and the mass ratio smaller than 0.7.
The absence of any evidence of the companion in the cross-correlation function 
together with the size of the radial velocity amplitude
also indicate a mass ratio of at most 0.7, and likely smaller. 
Furthermore, 
     the new data exclude the possibility that the companion has a mass in the planetary regime ($\leq$13\,$\mjup$).
     We show that the companion contributes significantly to the total luminosity of the system: 
     model-dependent estimates provide a minimum for the luminosity ratio $L_2/L_1$ of 0.2.
     \chaha8 is the fourth known spectroscopic brown dwarf or very low-mass stellar binary
     with determined orbital parameters, and the second known very young one.
     With an age of only about 3\,Myr, it is of particular interest to very low-mass formation and evolution theories.
     In contrast to most other spectroscopic binaries, it has a relatively long orbital period and 
     it might be possible to determine the astrometric orbit of the primary
     and, thus, the orbital inclination.
}


   \maketitle
%

\section{Introduction}
\label{sect:intro}

Brown dwarf (BD) and very low-mass stellar (VLMS, $M \leq 0.1\,\msun$) 
binaries are key objects to 
understand formation and evolution in the low-mass regime, in particular when they are very young.
Roughly a hundred BD/VLMS binaries have been detected in recent years,
predominantly by direct (adaptive optics or \emph{Hubble Space Telescope}) imaging 
(e.g., http://www.vlmbinaries.org). 
For several of these visual BD/VLM binaries, it
was possible to directly monitor the orbital motion and derive constraints on 
the orbital parameters and mass 
(e.g., Bouy et al. 2004; Dupuy et al. 2009;
Konopacky et al. 2010; Stumpf et al. 2010).
High-resolution spectroscopic monitoring for radial velocity (RV) variations
led to the detection of a few very close ($\lesssim$1\,AU)
spectroscopic BD/VLM binaries. For four of them, RV orbital parameters were determined:
\emph{PPl\,15}, a double-lined spectroscopic binary (SB2) in the Pleiades
(Basri \& Mart\'\i n 1999);
\emph{2MASS~J05352184-0546085} (hereafter 2M0535-05), a very young eclipsing SB2 system in Orion
(Stassun, Mathieu \& Valenti 2006);
\emph{\chaha8}, a very young spectroscopic binary in Chamaeleon\,I (Joergens \& M\"uller 2007);
and \emph{2MASS~J03202839-0446358}, an SB1 at the M/L transition (Blake et al. 2008).
Furthermore, for a few close visual BD/VLM binaries follow-up RV measurements permitted 
additional constraints of
their orbits, e.g., 
\emph{GJ\,569B} (Zapatero Osorio et al. 2004; Simon, Bender \& Prato 2006;
Konopacky et al. 2010); and \emph{2MASS~J07464256+2000321} (Konopacky et al. 2010).
When combined with angular distance measurements or eclipse detections,
spectroscopic binaries allow valuable dynamical mass determinations.
Either the astrometric orbit or the eclipse light curve can provide the
inclination $i$ of the orbital plane, which is the missing parameter in the spectroscopic mass $M_2 \sin i$.
The mass is the most important input parameter for evolutionary models, 
which at ages $<$10\,Myr and masses $\leq$0.3\,M$_\mathrm{\odot}$ rely on only three
measurements, the two components of the young eclipsing BD binary
2M0535-05 (Stassun et al. 2006; Mathieu et al. 2007) and the low-mass 
component of the pre-main sequence star UZ\,Tau\,E
(0.294$\pm$0.027\,M$_\mathrm{\odot}$; Prato et al. 2002).

After 2M0535-05,
\chaha8\footnote{Simbad name: [NC98]~Cha~HA~8} is
the second known very young ($\sim$3\,Myr) BD/VLM spectroscopic binary and, therefore,
of particular interest in the context of BD/VLM evolution and formation.
Cha\,H$\alpha$\,8 is a very low-mass member (M5.75-M6.5, Comer\'on et al. 2000; Luhman 2004, 2007)
of the nearby ($\sim$160\,pc) Cha\,I star-forming region. 
An estimate of its mass based on evolutionary model tracks 
(Baraffe et al. 1998) yields 0.07\,$\msun$ when using the values 
for effective temperature and bolometric luminosity by Comer\'on et al. (2000) and 0.10\,$\msun$
when using those by Luhman (2007), respectively.
Thus, \chaha8 is either a BD or a VLMS. 
A more detailed description of \chaha8 and its properties can be found in
Joergens \& M\"uller (2007).

RV data obtained until 2007 (Joergens \& M\"uller 2007) showed that \chaha8
has a companion in a few years orbit with a potentially very low mass
close to or in the planetary mass regime. This made
\chaha8 a candidate for being the first BD/VLMS with an RV planet.
We present here new RV measurements for \chaha8 
which significantly improve the RV orbit for the system.

\begin{table}
\begin{minipage}[t]{\columnwidth}
\begin{center}
\caption{
\label{tab:rvs} 
RV measurements of Cha\,H$\alpha$\,8.
}
\renewcommand{\footnoterule}{}  
\begin{tabular}{llll}
\hline
\hline
\myrule
Date      & HJD           & RV     & ~$\sigma_{RV}$\\
          &               & [km\,s$^{-1}$] & [km\,s$^{-1}$] \\
\hline
\myrule
2000 04 05  &    2451639.61095  &    14.591  & 0.400   \\ 
2000 04 24  &    2451658.72597 	&    15.177  & 0.400   \\ 
2002 03 06  &    2452339.68965 	&    17.499 \tablefootmark{a}  & 0.193 \\ 
2002 03 22  &    2452355.65264 	&    17.355 \tablefootmark{a}  & 0.150 \\ 
2002 04 16  &    2452380.61646 	&    17.651 \tablefootmark{a}  & 0.150 \\ %
2002 04 19  &    2452383.57565 	&    17.451 \tablefootmark{a}  & 0.379 \\ %
2005 03 21  &    2453450.62080 	&    14.791 \tablefootmark{a}  & 0.150 \\ 
2006 04 10  &    2453835.65109 	&    16.035  & 0.150   \\ 
2006 07 09  &    2453926.50137 	&    16.403  & 0.150 \\ 
2007 03 15  &    2454174.66101 	&    17.358 \tablefootmark{a}  & 0.402 \\ 
2007 03 22  &    2454181.69756 	&    17.563 \tablefootmark{a}  & 0.333 \\ 
\hline
\myrule
2007 06 08   &   2454260.49431  &  17.449 \tablefootmark{a} &  0.182 \\	
2007 07 21   &   2454302.54665  &  17.572 \tablefootmark{a} &  0.150 \\	
2008 01 02   &   2454467.82631  &  17.573 \tablefootmark{a} &  0.150 \\	
2008 07 09   &   2454657.48163  &  18.190 \tablefootmark{a} &  0.150 \\	
2009 01 02   &   2454833.80683  &  18.780 \tablefootmark{a} &  0.155 \\   
2010 01 09   &   2455205.74852  &  13.900 \tablefootmark{a} &  0.150 \\   
2010 03 01   &   2455257.26535  &  14.533 \tablefootmark{a} &  0.203 \\	

\hline
\end{tabular}
\tablefoot{Listed are new RV data (bottom panel) and 
previous RV measurements (top panel), which have been re-processed for this work.
HJD is given at the middle of the exposure; 
$\sigma_{RV}$ is the estimated error in the relative RVs.
An additional error of about 400\,m\,s$^{-1}$ has to be taken into account for the absolute RVs.}
\tablefoottext{a}{RV value is the average of two (three in the case of 2010 03 01) single consecutive measurements. 
}
\end{center}
\end{minipage}
\end{table}


\section{Radial velocities and orbital solution}
\label{sect:rvs}

\begin{figure}[t]
\centering
\includegraphics[width=\linewidth,clip]{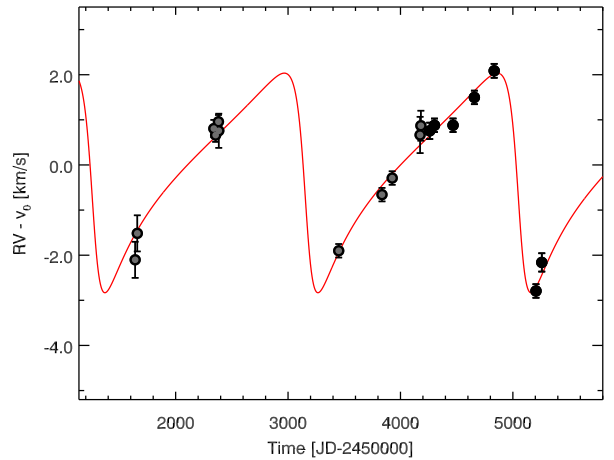} 
\caption{
\label{fig:orbit}
RV measurements of \chaha8 between 2000 and 2010 based on UVES/VLT spectra.
New RV data measured between Jun 2007 and Mar 2010 are marked with black-filled points,
previously obtained RV data with grey-filled points.
Overplotted is the best-fit Keplerian orbit, which has a semi-amplitude of 
2.4\,km\,s$^{-1}$, a period of 5.2\,years and an eccentricity of $e$=0.59.
}
\end{figure}

Spectroscopic observations of \chaha8 were carried out between 2000 and 2010 with
the \emph{Ultraviolet and Visual Echelle Spectrograph} (UVES, Dekker et al. 2000) 
attached to the VLT 8.2\,m KUEYEN telescope
at a spectral resolution $\lambda$/$\Delta \lambda$ of 40\,000 in the 
red optical wavelength regime.
RVs were measured from these spectra based on a cross-correlation technique
employing telluric lines for the wavelength calibration. 
Details of the data analysis can be found in Joergens (2006; 2008).

RV measurements for spectra taken between 2000 and 2002 
provided already evidence of an RV companion to \chaha8 (Joergens 2006).
Follow-up RV monitoring between 2005 and March 2007 allowed the determination of a 
first spectroscopic orbital solution for \chaha8 (Joergens \& M\"uller 2007).
Here, we present new RV data based on UVES spectra taken between June 2007 and March 2010. 
Table\,\ref{tab:rvs} lists the new RV measurements of \chaha8 
and previous RV data. The latter were reprocessed for this work
by using a slightly refined wavelength range for the cross-correlation to exclude
a region that had been found to be contaminated by telluric lines 
and imperfectly corrected CCD cosmetic blemishes and that introduced additional scatter.
The given errors $\sigma_{RV}$ are in most cases the sample standard deviation of 
two consecutive individual measurements
taking into account an additional minimum error of 150\,m\,s$^{-1}$.
This minimum error accounts for both RV noise due to activity and
fluctuations in the standard deviation for a small number of measurements.
It is derived from an estimate of the RV noise based on 
the night-to-night rms scatter in the data.

The combined data set spans 9.9 years of RV monitoring for \chaha8 between 2000 and 2010. 
We calculate an improved orbital solution based on these data by fitting the six free parameters
period, periastron time, eccentricity, longitude of periastron,
RV semi-amplitude, and system velocity. The fitting procedure uses 
a global optimization method that is based on the genetic algorithm PIKAIA (Charbonneau 2002), as 
in the previous work.
The reduced $\chi^2$ of the orbital fit is 1.45.

The fitted Kepler orbit is that of a companion revolving \chaha8 with a period of 1895\,d (5.2\,yr) 
on an eccentric ($e$=0.59) orbit and causing an RV semi-amplitude of 2.4\,km\,s$^{-1}$. 
The derived lower limit to the semi-major axis of the primary $a_1 \sin i$ is 0.34\,AU.
Figure\,\ref{fig:orbit} shows the RV measurements
and the RV curve of the best-fit Kepler model.
The complete list of determined orbital elements is given in Table\,\ref{tab:orbitparam}.

The minimum mass of the companion $M_2 \sin i$ in the case of a 
single-lined RV orbit depends on the primary mass, which is
not very precisely determined for \chaha8, as is common in this mass and age regime. 
Using the two available estimates for the primary mass
(0.07 and 0.10\,$\msun$, cf. Sect.\,\ref{sect:intro}),  
the mass $M_2 \sin i$ of the companion is
determined to be 25 and 31$\mjup$, respectively. 
The given errors in $M_2 \sin i$ (cf. Table\,\ref{tab:orbitparam})
are based on the fit and do
not take into account additional possible errors in the primary mass,
such as those introduced by the evolutionary models used to estimate the primary mass.
The semi-major axis of the companion, which also depends on the primary mass estimate,
is of the order of 1\,AU. 

The new orbit solution for \chaha8 is
based on more epochs of RV measurements that cover a significantly larger portion of the orbital phase
than the one derived by Joergens \& M\"uller (2007). 
As illustrated in Fig.\,\ref{fig:orbit}, 
previous data covered only about half of the total orbit, while
the new measurements
monitor the system at orbital phases that were so far completely unconstrained.
Thus, the constraints on the orbit are now much stronger than in 2007 (one should not be misled by
the apparently superior goodness of the fit of the previous orbit, which 
is in fact a result of undersampling).
Because of this much larger phase coverage, the refinement of the RV data,
and the introduction of a minimum error, 
an orbital solution is derived that has a
significantly longer orbital period, 
and a slightly larger RV semi-amplitude and eccentricity than previously anticipated. 
The newly derived companion mass is consequently higher 
but being within 1\,$\sigma$ still consistent with that found previously
(16$\pm$4\,$\mjup$ and 20$\pm$5\,$\mjup$; note that the error ranges for $M_2 \sin i$
are quoted incorrectly in Joergens \& M\"uller 2007).

It has already been demonstrated in detail (Joergens \& M\"uller 2007) that the detected RV variations
cannot be caused by activity. The main arguments are summarized in the following: 
(i) The timescale of the RV variations
is much too long to be explained by any phenomena that causes rotational modulation (the rotation rate
of \chaha8 is of the order of
a few days; Joergens \& Guenther 2001; Joergens et al. 2003). (ii) Accretion can also be excluded
as source of RV noise because neither mid-IR (Persi et al. 2000; Comer\'on et al. 2000)
nor $L$-band observations (Jayawardhana et al. 2003; Luhman et al. 2008)
indicate the presence of disk material surrounding \chaha8.
H$\alpha$ measurements
also provide no evidence of accretion (e.g. Mohanty et al. 2005). 

\begin{table}[t]
\begin{center}
\caption{
\label{tab:orbitparam} 
Orbital and physical parameters derived for the best-fit Keplerian model
of Cha\,H$\alpha$\,8.
}
\begin{tabular}{lc}
\hline
\hline
\\

Parameter  & Value \\

\hline
\myrule

$P$ (days)                \dotfill  & 1895   $\pm$ 132      \\
$T$ (HJD-2450000)         \dotfill  & 3163   $\pm$ 214      \\
$e$                       \dotfill  & 0.59   $\pm$ 0.22     \\
$\omega$ ($^{\circ}$)            \dotfill  & 106    $\pm$ 26       \\
$K$ (km\,s$^{-1}$)        \dotfill  & 2.433  $\pm$ 0.44     \\
$V_0$ (km\,s$^{-1}$)        \dotfill  & 16.693 $\pm$ 0.75     \\
\hline
$f(m)$ ($10^{-3}\,\msun$) \dotfill  & 1.488                \\
$M_2 \sin i $ ($\mjup$)   \dotfill  & 25 $\pm$ 7 \tablefootmark{a}, 31 $\pm$ 8 \tablefootmark{a} \\
$a_1 \sin i$ (AU)         \dotfill  & 0.34   $\pm$ 0.09     \\
$a_2 $ (AU)               \dotfill  & 1.02 $\pm$ 0.06 \tablefootmark{a}, 1.17 $\pm$ 0.07 \tablefootmark{a} \\

\hline

$N_{\rm meas}$           \dotfill   & 18    \\
Span (days)              \dotfill   & 3618  \\ 
$\sigma$ (O-C) (m/s)     \dotfill   & 228   \\
$\chi^{2}_{\rm red}$     \dotfill   & 1.45  \\

\hline
\end{tabular}
\tablefoot{
The given parameters are: orbital period,  
periastron time, eccentricity, longitude of periastron,
RV semi-amplitude, system velocity, 
mass function, lower limit of the companion mass,
lower limit of the semi-major axis of the primary,	  
semi-major axis of the companion, number of measurements,
time span of the observations, residuals, reduced $\chi^2$. 
}
\tablefoottext{a}{Derived parameter based on two available estimates 
for the primary mass of 0.07\,$\msun$ and 0.10\,$\msun$.
No further errors of the primary mass, e.g. as introduced by the use of
evolutionary models, have been taken into account here.}
\end{center}
\end{table}


\section{Discussion and conclusions}
\label{sect:concl}

Following up on the discovery of an RV companion orbiting the BD candidate \chaha8 
(Joergens \& M\"uller 2007), 
we have monitored the system between 2007 and 2010 with high-resolution spectroscopy 
with UVES at the VLT. 
The combined data set spans 9.9 years of RV monitoring of \chaha8 and 
constitutes the longest RV monitoring program of a BD or VLMS.
We have presented a significantly improved RV orbit solution of the system.
The best-fit Kepler orbit of \chaha8 has a 5.2\,yr period,
an RV semi-amplitude of 2.4\,km\,s$^{-1}$, and an eccentricity of 0.59.
We determined the spectroscopic minimum mass $M_2 \sin i$ for the companion 
to be 25$\pm$7\,$\mjup$ and 31$\pm$8\,$\mjup$ 
for the two available model-dependent mass estimates for the primary of 0.07\,$\msun$ and 0.10\,$\msun$,
respectively.

Based on the new RV data and orbit solution, we have been able to exclude 
that the companion of \chaha8 has a mass in the planetary mass regime ($\leq13\,\mjup$). 
New RV measurements (black-filled points in Fig.\,\ref{fig:orbit}) significantly improve the 
phase coverage compared to previously available data (grey-filled points in Fig.\,\ref{fig:orbit})
and show that the RV amplitude is larger than previously anticipated and that the orbit is inconsistent
with a planetary mass of the companion.

The companion of \chaha8 is with high probability of substellar nature.
This follows based on statistical considerations, as described in this paragraph,
and it follows also independently based on the absence of companion lines, as described later.
While RV orbits permit only the determination of lower mass limits $M_2 \sin i$
due to the unknown inclination $i$, for orbits oriented randomly in space,
there is a high probability 
that the true mass lies not too far from this limit.
In the following, we calculate the absolute companion mass $M_2$ for different inclinations.
For this purpose, $M_2$ is derived directly from the mass function rather than
using the value of $M_2 \sin i$ 
given in Table\,\ref{tab:orbitparam}, which is for non-negligible companion masses 
valid only for large inclinations.
Furthermore, it is taken into account that 
the companion contributes increasingly to the total luminosity when considering increasingly
smaller inclinations 
(see paragraph after next)
and, therefore, that the primary mass becomes increasingly low compared to the value 
derived for the unresolved luminosity. 
Here, the luminosity determined by Luhman (2007) for the unresolved system is used, 
which corresponds to a mass of the unresolved source of 0.10\,$\msun$.
We find that with a probability of greater than 50\%
(inclination greater than 60$^{\circ}$),
the companion of \chaha8 has a mass between 30 and 35\,$\mjup$.
With a probability higher than 87\% ($i \geq$ 30$^{\circ}$), it has a mass
between 30 and 69\,$\mjup$. 

The mass ratio $q\equiv M_2/M_1$ of \chaha8 seems to be on the small side compared 
to other known BD/VLM binaries:
it is at minimum 0.3, with more than 50\% probability smaller than 0.4, and 
with more than 87\% probability smaller than 0.7.
This is smaller than the mass ratio of the majority of BD/VLMS detected through
direct imaging, where 68\% have q$\geq$0.8 (value derived for binaries with a 
total mass $<$0.2$\msun$ as listed in the VLM binary archive as of July 2009 and applying small updates; 
cf. also Burgasser et al. 2007).

The companion of \chaha8 contributes significantly to the total luminosity
of the system.
The luminosity ratio $L_2 / L_1$ is estimated for different inclinations
by employing low-mass evolutionary model tracks (Baraffe et al. 1998) and assuming the coevality of the
binary components. 
For the luminosity of the unresolved system and its effective temperature, 
the values determined by Luhman (2007; $L_{\rm{tot}}$=0.037\,$\lsun$, $T_{\rm{eff}}$=3024\,K) are used,
which correspond to a mass of the unresolved source of 0.10\,$\msun$.
At minimum, i.e. for $i$=90$^{\circ}$, the luminosity ratio $L_2 / L_1$ of the \chaha8
system is found to be 0.2.
For larger inclinations, it becomes increasingly large, e.g., $L_2 / L_1$=0.3 for $i$=60$^{\circ}$ and 
$L_2 / L_1$=0.7 for $i$=30$^{\circ}$, respectively.
The limiting case is to consider a companion that is as luminous as the primary
(i.e., $L_1$=$L_2$=$L_{\rm{tot}}/2$). 
Comparison with evolutionary tracks shows that in this case
the system would be slightly older (almost 5\,Myr) and both components would have a mass of $M_1$=$M_2\approx$0.085\,$\msun$. 
This constitutes the model-dependent minimum mass of the primary when using the parameters by Luhman (2007). Furthermore, it
implies a minimum orbital inclination of 24$^{\circ}$.
However, we show in the next paragraph that the components cannot 
have equal luminosities and masses but that 
$L_2/L_1$ and q are instead significantly smaller than 1.

There is no evidence of spectral lines of the companion in the spectra: neither exhibits
the cross-correlation function a second peak, nor does its width or shape (bisector velocity span)
correlate with RV. 
This implies that the companion is either too faint to be directly detected
and/or that its signal is engulfed completely by the rotational broadening of the line profile
($v \sin i$=16$\pm$3\,km\,s$^{-1}$; Joergens \& Guenther 2001).
In the latter case, blending effects can cause the net RV signal to be weaker than the true RV amplitude.
To determine the maximum mass and luminosity ratio that is consistent with the absence of secondary lines
and at the same time with the measured RV variation, we simulate rotationally broadened Gaussian functions representing the 
two components of a hypothetical binary using the eccentricity and period as determined here,
the primary mass of \chaha8, allowing RV amplitudes of $\geq$2.4\,km\,s$^{-1}$, 
inclinations of 24--90$^{\circ}$, and involving 
again evolutionary models to estimate the luminosity ratio.
The case in which both components
have equal masses, and, therefore, equal luminosities, 
can be excluded immediately because the net RV signal would then be zero.
More generally, we find that the mass ratio cannot be too high because the simulated RV curves
would be of too small net amplitude due to blending. We derive
a conservative upper limit to both\footnote{Note 
that for young BDs, the mass-luminosity relation according to the applied evolutionary tracks 
is much less steep than for old field BDs, leading to constraints on the
mass ratio that are of similar order as the luminosity ratio.} the mass and luminosity ratio of about 0.7.

Because of its very young age ($\sim$3\,Myr)
and the small number of known BD/VLMS spectroscopic binaries with determined orbital
parameters, the very low-mass binary system \chaha8 is 
potentially of great interest for understanding formation and evolution
close to and below the substellar borderline.
\chaha8 has a relatively long orbit for a spectroscopic system
and might therefore become one of the few spectroscopic binaries for which
the relative astrometric orbit of the primary can be determined and for which
absolute dynamical masses can be measured.
Therefore, an astrometric monitoring program of this object is ongoing.

\begin{acknowledgements}
We acknowledge the excellent work of the ESO staff at Paranal, 
who took the data presented here in service mode. Furthermore,
we are grateful to ESO for the possibility of immediate observations in 2010, 
which allowed us to sample the RV minimum of \chaha8.
We would also like to thank an anonymous referee for very helpful comments.
Part of this work was funded by the ESF.
This publication has made use of the Very-Low-Mass Binaries Archive housed 
at http://www.vlmbinaries.org and maintained by Nick Siegler, Chris Gelino, and Adam Burgasser.
\end{acknowledgements}


\begin{thebibliography}{}
\bibitem{BCAH} Baraffe, I., Chabrier, G., Allard, F. \& Hauschildt, P. H. 1998, A\&A 337, 403

\bibitem{basri1999} Basri, G., \& Mart\'{\i}n, E. L. 1999, ApJ 118, 2460 

\bibitem{blake2008} Blake, C. H., Charbonneau, D., White, R. J., Torres, G., Marley, M. S., Saumon D.
                    ApJ 678, L125 

\bibitem{bouy2004} Bouy, H., Duch\'ene, G., K\"ohler, R. et al. 2004, A\&A 423, 341-352

\bibitem{burgasserppv} Burgasser, A. J., Reid, I. N., Siegler, N., Close, L., Allen, P., Lowrance, P. \&
  Gizis J., 2007, 
  in Protostars and Planets\,V, ed. by B. Reipurth, D. Jewitt \& K. Keil 
(Tucson, University of Arizona Press), 427

\bibitem{charbonneau2002} Charbonneau, P., 2002, Release Notes for PIKAIA 1.2, 
                  NCAR Technical Note 451+STR (Boulder: National Center for Atmospheric Research)

\bibitem{comeron00} Comer\'on, F., Neuh\"auser, R., \& Kaas, A. A. 2000, A\&A, 
			359, 269 

\bibitem{dekker} Dekker, H., D'Odorico, S., Kaufer, A., Delabre, B., \& Kotzlowski, H. 2000, 
                 in SPIE Vol. 4008, ed. by M.Iye, A.Moorwood, 534

\bibitem{dupuy} Dupuy et al. 2009 ApJ 706, 328

\bibitem{rayjay} Jayawardhana, R., Ardila, D.R., Stelzer, B. \& Haisch Jr. K.E. 2003, AJ 126, 1515

\bibitem{joergens01} Joergens, V., \& Guenther, E. 2001, A\&A, 379, L9

\bibitem{joergens03} Joergens, V., Fern\'andez, M., Carpenter, J. M., \& Neuh\"auser, R. 2003, 
                     ApJ 594, 971

\bibitem{J06a} Joergens, V. 2006, A\&A 446, 1165

\bibitem{JM07} Joergens, V., \& M\"uller, A. 2007, ApJ 666, L113

\bibitem{J08} Joergens, V. 2008, A\&A 492, 545

\bibitem{konopacky2010} Konopacky et al. 2010, ApJ 711, 1087

\bibitem{luhman2004} Luhman, K. L. 2004, ApJ 614, 398

\bibitem{luhman2007} Luhman, K. L. 2007, ApJS 173, 104

\bibitem{luhmanetal2008} Luhman, K. L., Allen, L. E., Allen, P. R. et al.
                     2008, ApJ 675, 1375

\bibitem{mathieu2007} Mathieu, R. D., Baraffe, I., Simon, M., Stassun, K. G. \& White, R., 2007,
in Protostars and Planets\,V, ed. by B. Reipurth, D. Jewitt \& K. Keil,
(Tucson, University of Arizona Press), 411

\bibitem{mohanty2005} Mohanty, S., Jayawardhana, R. \& Basri G. 2005, ApJ 626, 498

\bibitem{prato2002} Prato, L., Simon, M., Mazeh, T., Zucker, S., McLean, I. S. 2002, ApJ 579 L99	

\bibitem{persi} Persi, P., et al. 2000, A\&A, 357, 219

\bibitem{simon2006} Simon, M., Bender, C. \& Prato, L. 2006, ApJ 644, 1183

\bibitem{stassun2006} Stassun, K. G., Mathieu, R. D. \& Valenti, J. A. 2006, Nature 440, 311

\bibitem{stumpf} Stumpf, M. B., Brandner, W., Henning, Th., Bouy, H., Koehler, R., Hormuth, F., Joergens, V., Kasper, M. 
2010, A\&A subm.

\bibitem{zapatero2004} Zapatero Osorio, M. R., Lane, B. F., Pavlenko, Ya., 
  Mart\'\i n E. L.,  Britton, M. \& Kulkarni, S. R. 2004, ApJ 615, 958

\end{thebibliography}
\end{document}